# Upgraded Approximation of Non-Binary Alphabets for Polar Code Construction

**Arash Ghayoori and T. Aaron Gulliver**

Department of Electrical and Computer Engineering,
University of Victoria,
Victoria, BC, Canada

agullive@ece.uvic.ca
arashgh@ece.uvic.ca

**Abstract** An algorithm is presented for approximating a single-user channel with a prime input alphabet size. The result is an upgraded version of the channel with a reduced output alphabet size. It is shown that this algorithm can be used to reduce the output alphabet size to the input alphabet size in most cases.

*I. Introduction*

Polar codes [1] were recently introduced by Arikan for binary-input, memoryless, symmetric-output (BMS) channels. The definition of polar codes was recently generalized to channels with a prime input alphabet size [2].

Although the definitions of polar codes presented in [1] and [2] are explicit, finding a simple and efficient means of constructing polar codes is still a challenge. A direct implementation of the definitions to construct polar codes is intractable because of the computational complexity. Thus, approximations need to be employed. Recently, Tal and Vardy [3] proposed a method for efficient construction of polar codes in the context of a single-user, binary-input channel, using quantization. They considered the fact that every channel is associated with a probability of error and used degrading and upgrading quantization to derive lower and upper bounds, respectively. The degrading quantization results in a channel degraded with respect to the original one, while the upgrading quantization results in a channel upgraded with respect to the original one. They then constructed a polar code based on the degraded channel while measuring the distance from the optimal result by considering the upgraded channel. Their method was analyzed in [4]. The degraded approximation proposed in [3] has recently been generalized to single-user channels and MACs with a prime input alphabet size [5]. The aim of this paper is to generalize the upgraded approximation proposed in [3] to single-user channels with prime input alphabet size. For the case of binary channels, it was shown in [3] that the upgraded channel by Tal and Vardy is very close to the original channel. In this paper, a similar approach is employed to upgrade channels with non-binary input alphabets. Therefore it can be expected that the resulting upgraded channel will be close to the original channel.

The remainder of this paper is structured as follows. The notation and some preliminary results are presented in Section II. In Section III, Lemmas 7 and 9 in [3] are extended from binary to ternary alphabets. The generalization to larger input alphabets is given in Section IV. Note that, unlike the binary case, merging symbols will not reduce the output alphabet size for non-binary input alphabets. However, in Section V, an algorithm is proposed which can reduce the output alphabet size to the size of the input alphabet. Section VI considers the possibility of the existence of odd symbols (symbols that have zero probability for at least one input) in the output alphabet. Solutions for several cases are provided. In Section VII, a general solution when an odd symbol exists in the output alphabet is given. It is shown that with this approach, one can simply handle all cases of odd symbols. This is achieved by modifying these symbols so that they can be considered as normal symbols, and then the results in the previous sections can applied. This generalizes the algorithm presented to all single-user channels. Finally, a summary of the main contributions is given in Section VIII.

## II. Preliminaries

Let $W: X \rightarrow Y$ be a general single-user channel, where $X = \{0, 1, 2, ..., p-1\}$ is the input alphabet, $p$ is prime, and $Y$ is the finite output alphabet. This channel is specified by the transition probabilities $W(y|x)$, which is the probability of observing $y \in Y$ when $x \in X$ is transmitted through $W$. The definition of an upgraded single-user channel with a non-binary input alphabet is similar to the binary case in [3].

**Definition 1:** For the channel $Q': X \rightarrow Z$ to be upgraded with respect to $W: X \rightarrow Y$, an intermediate channel $P: Z \rightarrow Y$, must be found such that for all $z \in Z$ and $x \in X$

$$W(y|x) = \sum_{z \in Z} Q'(z|x).P(y|z).$$

We write this as $Q' \geq W$ ($Q'$ is upgraded with respect to $W$).

The main intuition behind the work of Tal and Vardy [3] is now given. The purpose of merging output symbols is to reduce the output alphabet size so that it becomes manageable, and therefore easier to work with. However, merging symbols arbitrarily, i.e., merging symbols with different likelihood ratios, will result in a degraded channel, whereas merging symbols with the same likelihood ratios will result in an upgraded channel.

The likelihood ratio for non-binary alphabets is a vector, whereas for a binary alphabet, it is a scalar. The definition of the likelihood ratio ($LR$) used in this paper is as follows [6].

**Definition 2:** Let $x = (x_0, x_1, ..., x_{p-1})$ be a $p$-dimensional probability vector of real numbers such that $x_i \geq 0$ for all $i$ and $\sum_{i=0}^{p-1} x_i = 1$. The likelihood ratio values associated with this vector are defined as

$$w_i = x_0 / x_i$$
$$i = 0, 1, 2, \ldots, p-1$$

Note that for all $x$, $w_0 = 1$. Define the $LR$ vector representation of $x$ as the $p$-dimensional vector $w = (w_0, w_1, ..., w_{p-1})$.

In this paper, the output symbols are divided into 3 categories as follows.

**Definition 3:** A "normal" symbol is a symbol which can occur for all given inputs, i.e., the probability of occurrence is nonzero for any input. The norm of the likelihood ratio vector of a "normal" symbol is a real number which is greater than or equal to one.

**Definition 4:** An "odd" symbol is a symbol which cannot occur for at least one given input, i.e., the probability of occurrence is zero for at least one input. The norm of the likelihood ratio vector of an "odd" symbol is infinity.

***Definition 5:*** A "leftover" symbol is a symbol, which is either likely to happen for only one of the given inputs, i.e., its probability of occurrence is zero for all inputs except one, or cannot occur for only two inputs. Note that a "leftover" symbol is a special case of an "odd" symbol.

Using the approach given earlier for merging output symbols together, a degraded approximation for the non-binary case was introduced in [4]. Considering the approach in [3], an upgraded approximation for the non-binary case is presented here. For simplicity, we first consider the ternary case and then generalize the results to the case of prime input alphabet sizes.

### III. Upgraded Approximation

In this section, Lemmas 7 and 9 in [3] are generalized to channels with a prime input alphabet size. As mentioned previously, the lemmas are first given for ternary channels and then generalized to prime input alphabet sizes in the next section. Note that in the following lemmas, it is assumed that all output symbols are normal, i.e., they can occur for all possible inputs. Sections VI and VII consider the case of odd symbols in the output alphabet and a general solution is provided.

***Lemma 1:*** Let $W: X \to Y$ be a ternary single-user channel and let $y_1$ and $y_2$ be symbols in the output alphabet $Y$. Denote

$$\lambda_1 = LR(y_1) = (\lambda_{10}, \lambda_{11}, \lambda_{12}) = \left(1, \frac{a_1}{b_1}, \frac{a_1}{c_1}\right),$$

and

$$\lambda_2 = LR(y_2) = (\lambda_{20}, \lambda_{21}, \lambda_{22}) = \left(1, \frac{a_2}{b_2}, \frac{a_2}{c_2}\right).$$

Assume that $1 \leq \|\lambda_1\| \leq \|\lambda_2\|$. Further, let $a_1 = W(y_1 | 0)$, $b_1 = W(y_1 | 1)$, and $c_1 = W(y_1 | 2)$ and assume that $W(y_1 | 0) = W(y_2 | 0)$, $W(y_1 | 1) = W(y_2 | 1)$ and $W(y_1 | 2) = W(y_2 | 2)$. Define $\alpha_2, \beta_2$ and $\gamma_2$ as follows

$$\alpha_2 = \frac{\lambda_{22}\lambda_{21}(a_1 + b_1 + c_1)}{\lambda_{22}\lambda_{21} + \lambda_{22} + \lambda_{21}}$$

$$\beta_2 = \frac{\lambda_{22}(a_1 + b_1 + c_1)}{\lambda_{22}\lambda_{21} + \lambda_{22} + \lambda_{21}},$$

$$\gamma_2 = \frac{\lambda_{21}(a_1 + b_1 + c_1)}{\lambda_{22}\lambda_{21} + \lambda_{22} + \lambda_{21}}$$

and for real numbers $\alpha, \beta, \gamma$ and $x \in X$, define

$$t(\alpha, \beta, \gamma \mid x) = \begin{cases} \alpha, x = 0 \\ \beta, x = 1 \\ \gamma, x = 2 \end{cases}.$$

The output alphabet $Z$ is given by

$$Z = Y \setminus \{y_1, y_2\} \cup \{z_2\}.$$

For all $x \in X$ and $z \in Z$

$$Q'(z \mid x) = \begin{cases} W(y_2 \mid x) + t(\alpha_2, \beta_2, \gamma_2 \mid x), z = z_2 \\ W(z \mid x), z \neq z_2 \end{cases}.$$

Then $Q' \geq W$, that is, $Q'$ is upgraded with respect to $W$.

*Proof:* Define the intermediate channel $P : Z \to Y$ as follows

$$P = (p_1 \quad p_2)$$
$$p_1 = \frac{1}{2}$$
$$p_2 = \frac{1}{2}.$$

$P$ satisfies the requirements given in Definition 1, and therefore, $Q' \geq W$.

∎

Note that Lemma 1 can only be used to upgrade the channel when the two symbols being merged have the same probabilities of occurrence for given inputs, as this condition is required in the proof. For this reason, in the proposed algorithm, the following lemma is employed.

*Lemma 2:* Let $W : X \to Y$ be a ternary single-user channel and let $y_1, y_2$ and $y_3$ be symbols in the output alphabet $Y$. Denote

$$\lambda_1 = LR(y_1) = (\lambda_{10}, \lambda_{11}, \lambda_{12}) = \left(1, \frac{a_1}{b_1}, \frac{a_1}{c_1}\right),$$

$$\lambda_2 = LR(y_2) = (\lambda_{20}, \lambda_{21}, \lambda_{22}) = \left(1, \frac{a_2}{b_2}, \frac{a_2}{c_2}\right)$$

and

$$\lambda_3 = LR(y_3) = (\lambda_{30}, \lambda_{31}, \lambda_{32}) = \left(1, \frac{a_3}{b_3}, \frac{a_3}{c_3}\right).$$

Assume that $1 \leq \|\lambda_1\| \leq \|\lambda_2\| \leq \|\lambda_3\|$.

Next, let $a_2 = W(y_2 | 0), b_2 = W(y_2 | 1)$ and $c_2 = W(y_2 | 2)$, and define $\alpha_1, \alpha_2, \alpha_3, \beta_1, \beta_2, \beta_3, \gamma_1, \gamma_2$ and $\gamma_3$ as follows

$$\alpha_3 = \frac{\lambda_{32}\lambda_{31}(c_2\lambda_{12} - b_2\lambda_{11})}{\lambda_{12}\lambda_{31} - \lambda_{11}\lambda_{32}}$$

$$\alpha_1 = \frac{\lambda_{11}\lambda_{12}(b_2\lambda_{31} - c_2\lambda_{32})}{\lambda_{12}\lambda_{31} - \lambda_{11}\lambda_{32}}$$

$$\beta_3 = \frac{\lambda_{32}(c_2\lambda_{12} - b_2\lambda_{11})}{\lambda_{12}\lambda_{31} - \lambda_{11}\lambda_{32}}$$

$$\beta_1 = \frac{\lambda_{12}(b_2\lambda_{31} - c_2\lambda_{32})}{\lambda_{12}\lambda_{31} - \lambda_{11}\lambda_{32}}$$

$$\gamma_3 = \frac{\lambda_{31}(c_2\lambda_{12} - b_2\lambda_{11})}{\lambda_{12}\lambda_{31} - \lambda_{11}\lambda_{32}}$$

$$\gamma_1 = \frac{\lambda_{11}(b_2\lambda_{31} - c_2\lambda_{32})}{\lambda_{12}\lambda_{31} - \lambda_{11}\lambda_{32}}$$

$$\beta_2 = \gamma_2 = 0$$

$$\alpha_2 = \frac{a_2(\lambda_{12}\lambda_{31} - \lambda_{11}\lambda_{32}) - b_2\lambda_{11}\lambda_{31}(\lambda_{12} - \lambda_{32}) - c_2\lambda_{12}\lambda_{32}(\lambda_{11} + \lambda_{31})}{\lambda_{12}\lambda_{31} - \lambda_{11}\lambda_{32}}$$

Let $t(\alpha, \beta, \gamma)$ be as in Lemma 1 and define the channel $Q': X \to Z$ as follows.

The output alphabet $Z$ is given by

$$Z = Y \setminus \{y_1, y_2, y_3\} \cup \{z_1, z_2, z_3\}.$$

For all $x \in X$ and $z \in Z$

$$Q'(z | x) = \begin{cases} W(y_1 | x) + t(\alpha_1, \beta_1, \gamma_1 | x), z = z_1 \\ W(y_3 | x) + t(\alpha_3, \beta_3, \gamma_3 | x), z = z_3 \\ t(\alpha_2, \beta_2, \gamma_2 | x), z = z_2 \\ W(z | x), z \neq z_1, z_2, z_3 \end{cases}.$$

Then $Q' \geq W$, that is, $Q'$ is upgraded with respect to $W$.

*Proof:* Define the intermediate channel $P: Z \to Y$ as follows

$$P = \begin{pmatrix} p_1 & q_1 & 0 \\ 0 & 1 & 0 \\ 0 & q_2 & p_2 \end{pmatrix}$$

with

$$p_1 = \frac{a_1}{a_1 + \alpha_1} = \frac{b_1}{b_1 + \beta_1} = \frac{c_1}{c_1 + \gamma_1}$$

$$q_1 = \frac{\alpha_1}{a_1 + \alpha_1} = \frac{\beta_1}{b_1 + \beta_1} = \frac{\gamma_1}{c_1 + \gamma_1}$$

$$q_2 = \frac{\alpha_3}{a_3 + \alpha_3} = \frac{\beta_3}{b_3 + \beta_3} = \frac{\gamma_3}{c_3 + \gamma_3}$$

$$p_2 = \frac{a_3}{a_3 + \alpha_3} = \frac{b_3}{b_3 + \beta_3} = \frac{c_3}{c_3 + \gamma_3}$$

$P$ satisfies the requirements given in Definition 1, and therefore, $Q' \geq W$.

∎

Note that Lemma 2 transforms 3 output symbols into 3 new symbols which upgrades the channel but does not reduce the output alphabet size. Later, an algorithm will be presented which merges symbols and thus reduces this size.

*Proposition 1:* Let $A = \begin{pmatrix} a \\ 0 \\ 0 \end{pmatrix}$ and $B = \begin{pmatrix} b \\ 0 \\ 0 \end{pmatrix}$ be two symbols in the output alphabet of $W: X \to Y$, and let $W$ be as in Lemma 2. Merging these two symbols together will upgrade the channel.

*Proof:* Let $C = A \oplus B = \begin{pmatrix} a+b \\ 0 \\ 0 \end{pmatrix}$ be the result of merging A and B. Define the intermediate channel $P: Z \to Y$ as follows

$$P = \begin{pmatrix} p & q \end{pmatrix}$$
$$p = \frac{a}{a+b}$$
$$q = \frac{b}{a+b}.$$

$P$ satisfies the requirements given in Definition 1, and therefore the channel is upgraded.

∎

## IV. Generalization of Lemma 2 to a Prime Input Alphabet Size

This section considers the generalization of Lemma 2 to the case of single-user channels with prime input alphabet sizes. This requires an explanation of the intuition behind the results in the previous section. In Lemma 2, the 3 symbols were first arranged in ascending order based on the norm of their likelihood ratio vectors. Then, the symbol in the middle was divided into three new symbols, two with the same likelihood ratio vectors as the other two original (unchanged) symbols, and the remaining symbol which is a leftover symbol. This leftover symbol will have a nonzero value for only one of the given inputs. Following this approach, Lemma 2 can be extended to other prime input alphabet sizes. The only difference is that for input alphabet sizes greater than three, the leftover symbol will be zero for only two input symbols. Section VII contains an example that shows how this generalization is done for input alphabet size $p = 5$.

## V. Output Alphabet Size Reduction Algorithm

Recall that for channels with non-binary input alphabets, Lemma 2 will provide a channel which is upgraded with respect to the original channel, and Lemma 1 can be applied only when the values of the two symbols being merged are equal for given inputs. However, Lemma 2 does not reduce the output alphabet size which is the main objective of merging symbols. The following algorithm provides a solution to this problem. Note that this algorithm can be used to reduce the output alphabet size to the size of the input alphabet when all symbols are normal. The steps of this algorithm are given below.

***Step 1.*** Arrange the output symbols in ascending order based on the norm of their likelihood ratio vectors.

***Step 2.*** Merge the first 3 symbols in the new arrangement using Lemma 2. This will result in 3 new symbols, one of which is a leftover symbol.

***Step 3.*** Move the leftover symbol to the right end of the arrangement (the leftover symbol is considered to have an infinite likelihood ratio, thus it is moved to the end because it has the highest likelihood ratio).

***Step 4.*** Repeat Steps 2 and 3 until two normal symbols remain, and the others are all leftover symbols. Keep the two normal symbols.

***Step 5.*** Remove the zero rows in the leftover symbols, and consider them as normal output symbols. Repeat Steps 1 to 4 for these symbols.

***Step 6.*** Restore the zero rows to the two normal symbols remaining in Step 5 and keep them.

***Step 7.*** Continue Steps 5 and 6, until the leftover symbols are in the same form as the ternary case (once the zero rows have been removed). Use Proposition 1 to merge them and keep the resulting symbol.

***Step 8.*** Restore the zero rows to the resulting symbol in Step 7.

## *VI. Occurrence of "Odd" Symbols in the Output Alphabet*

We now consider the case of odd symbols in the output alphabet. Recall that an odd symbol is a symbol which cannot occur for at least one input. Therefore, the norm of its likelihood ratio vector is infinity. The existence of odd symbols in the output alphabet increases the complexity of merging symbols to reduce the output alphabet size while simultaneously upgrading the channel. However, solutions for several cases are provided in this section, while the general problem is addressed in the next section. For simplicity, the following lemma is proven for the ternary case, however it can easily be extended to all prime input alphabet sizes.

***Lemma 3:*** Let $W: X \to Y$ be a ternary single-user channel and let $y_1, y_2$ be two normal symbols and $y_3$ an odd symbol in the output alphabet $Y$. Denote

$$\lambda_1 = LR(y_1) = (\lambda_{10}, \lambda_{11}, \lambda_{12}) = \left(1, \frac{a_1}{b_1}, \frac{a_1}{c_1}\right)$$

and

$$\lambda_2 = LR(y_2) = (\lambda_{20}, \lambda_{21}, \lambda_{22}) = \left(1, \frac{a_2}{b_2}, \frac{a_2}{c_2}\right).$$

Assume that $1 \leq \|\lambda_1\| \leq \|\lambda_2\| < \infty$, and let

$a_2 = W(y_2|0), b_2 = W(y_2|1), c_2 = W(y_2|2), a_3 = W(y_3|0), b_3 = W(y_3|1)$ and $c_3 = W(y_3|2)$.

Further, assume that $b_3 = 0$ and define $\alpha_1, \beta_1, \gamma_1, \alpha_3, \beta_3$ and $\gamma_3$ as follows

$$\beta_1 = b_2, \beta_3 = 0$$

$$\alpha_1 = \lambda_{11} b_2, \alpha_3 = a_2 - \lambda_{11} b_2$$

$$\gamma_1 = \frac{\lambda_{11} b_2}{\lambda_{12}}, \gamma_3 = \frac{\lambda_{12} c_2 - \lambda_{11} b_2}{\lambda_{12}}$$

Let $t(\alpha, \beta, \gamma)$ be as in Lemma 1 and define the channel $Q': X \to Z$ as follows.

The output alphabet $Z$ is given by

$$Z = Y \setminus \{y_1, y_2, y_3\} \cup \{z_1, z_3\}.$$

For all $x \in X$ and $z \in Z$

$$Q'(z|x) = \begin{cases} W(y_1|x) + t(\alpha_1, \beta_1, \gamma_1|x), z = z_1 \\ W(y_3|x) + t(\alpha_3, \beta_3, \gamma_3|x), z = z_3 \\ W(z|x), z \neq z_1, z_3 \end{cases}.$$

Then $Q' \geq W$, that is, $Q'$ is upgraded with respect to $W$.

***Proof:*** Define the intermediate channel $P: Z \to Y$ as follows

$$P = \begin{pmatrix} p_1 & p_2 & 0 \\ 0 & q_2 & q_1 \end{pmatrix}$$

where

$$p_1 = \frac{a_1}{a_1 + \alpha_1} = \frac{b_1}{b_1 + \beta_1} = \frac{c_1}{c_1 + \gamma_1}$$

$$p_2 = \frac{\alpha_1}{a_1 + \alpha_1} = \frac{\beta_1}{b_1 + \beta_1} = \frac{\gamma_1}{c_1 + \gamma_1}$$

$$q_2 = \frac{\alpha_3}{a_3 + \alpha_3} = \frac{\gamma_3}{c_3 + \gamma_3}$$

$$q_1 = \frac{a_3}{a_3 + \alpha_3} = \frac{c_3}{c_3 + \gamma_3}$$

$P$ satisfies the requirements given in Definition 1, and therefore, $Q' \geq W$.

∎

Lemma 3 shows that if there is an odd symbol in the output alphabet which has zero probability for only one input, then it can be merged with two normal symbols. This results in two new symbols, one normal and one odd. This odd symbol has exactly the same form as the previous odd symbol (the zero probability occurs for the same input as the previous symbol). This lemma provides an upgraded channel with a reduced output alphabet size.

***Proposition 2:*** If there is a leftover symbol in the output alphabet which is only likely to happen for one given input, leave it unchanged until the end of the output alphabet size reduction algorithm, where you want to tackle the leftovers. Then, merge it with similar leftovers (leftovers that are only likely to happen for the same given input). This way, the channel is upgraded while the output alphabet size is being reduced at the same time.

The following lemma, considers the case when the odd symbol is unlikely to occur for two given inputs. For simplicity, this lemma is stated for input alphabet size $p = 5$. However, the generalization to other input alphabet sizes can be done easily.

***Lemma 4:*** Let $W: X \rightarrow Y$ be a single-user channel with input alphabet size $p = 5$ and let $y_1, y_2$ be two normal symbols and $y_3$, an odd symbol in the output alphabet $Y$. Denote

$$\lambda_1 = LR(y_1) = (\lambda_{10}, \lambda_{11}, \lambda_{12}, \lambda_{13}, \lambda_{14}) = \left(1, \frac{a_1}{b_1}, \frac{a_1}{c_1}, \frac{a_1}{d_1}, \frac{a_1}{e_1}\right)$$

and

$$\lambda_2 = LR(y_2) = (\lambda_{20}, \lambda_{21}, \lambda_{22}, \lambda_{23}, \lambda_{24}) = \left(1, \frac{a_2}{b_2}, \frac{a_2}{c_2}, \frac{a_2}{d_2}, \frac{a_2}{e_2}\right).$$

Assume that $1 \leq \|\lambda_1\| \leq \|\lambda_2\| < \infty$, and let

$a_2 = W(y_2 | 0), b_2 = W(y_2 | 1), c_2 = W(y_2 | 2), d_2 = W(y_2 | 3), e_2 = W(y_2 | 4), a_3 = W(y_3 | 0),$
$b_3 = W(y_3 | 1), c_3 = W(y_3 | 2), d_3 = W(y_3 | 3)$ and $e_3 = W(y_3 | 4)$.

Assume that $b_3 = c_3 = 0$ and define $A_1, B_1, C_1, D_1, E_1, A_2, B_2, C_2, D_2, E_2, A_3, B_3, C_3, D_3$ and $E_3$ as follows

$A_1 = \lambda_{11} b_2, A_2 = 0, A_3 = a_2 - \lambda_{11} b_2$

$B_1 = b_2, B_2 = 0, B_3 = 0$

$C_1 = \frac{\lambda_{11} b_2}{\lambda_{12}}, C_2 = \frac{\lambda_{12} c_2 - \lambda_{11} b_2}{\lambda_{12}}, C_3 = 0$

$D_1 = \frac{\lambda_{11} b_2}{\lambda_{13}}, D_2 = 0, D_3 = \frac{\lambda_{13} d_2 - \lambda_{11} b_2}{\lambda_{13}}$

$E_1 = \frac{\lambda_{11} b_2}{\lambda_{14}}, E_2 = 0, E_3 = \frac{\lambda_{14} c_2 - \lambda_{11} b_2}{\lambda_{14}},$

and for real numbers $A, B, C, D, E$ and $x \in X$, define

$$t(A,B,C,D,E \mid x) = \begin{cases} A, x=0 \\ B, x=1 \\ C, x=2 \\ D, x=3 \\ E, x=4 \end{cases}.$$

The output alphabet $Z$ is given by

$$Z = Y \setminus \{y_1, y_2, y_3\} \cup \{z_1, z_2, z_3\}.$$

For all $x \in X$ and $z \in Z$

$$Q'(z \mid x) = \begin{cases} W(y_1 \mid x) + t(A_1, B_1, C_1, D_1, E_1 \mid x), z = z_1 \\ W(y_3 \mid x) + t(A_3, B_3, C_3, D_3, E_3 \mid x), z = z_3 \\ t(A_2, B_2, C_2, D_2, E_2 \mid x), z = z_2 \\ W(z \mid x), z \neq z_1, z_2, z_3 \end{cases}.$$

Then $Q' \geq W$, that is, $Q'$ is upgraded with respect to $W$.

*Proof:* Define the intermediate channel $P: Z \to Y$ as follows

$$P = \begin{pmatrix} p_1 & q_1 & 0 \\ 0 & 1 & 0 \\ 0 & q_2 & p_2 \end{pmatrix}$$

$$p_1 = \frac{a_1}{a_1 + A_1} = \frac{b_1}{b_1 + B_1} = \frac{c_1}{c_1 + C_1} = \frac{d_1}{d_1 + D_1} = \frac{e_1}{e_1 + E_1}$$

$$q_1 = \frac{A_1}{a_1 + A_1} = \frac{B_1}{b_1 + B_1} = \frac{C_1}{c_1 + C_1} = \frac{D_1}{d_1 + D_1} = \frac{E_1}{e_1 + E_1}$$

$$q_2 = \frac{A_3}{a_3 + A_3} = \frac{D_3}{d_3 + D_3} = \frac{E_3}{e_3 + E_3}$$

$$p_2 = \frac{a_3}{a_3 + A_3} = \frac{d_3}{d_3 + D_3} = \frac{e_3}{e_3 + E_3}.$$

$P$ satisfies the requirements given in Definition 1, and therefore, $Q' \geq W$.

∎

Lemma 4 shows that if there is an odd symbol in the output alphabet which has zero probability for only two inputs, it can be merged with two normal symbols to obtain three new symbols. These symbols are a normal symbol, a leftover and an odd symbol with the same form as the previous odd symbol (the zero probabilities for the new odd symbol are the same as for the previous odd symbol). This lemma results in an upgraded channel, but the output alphabet size has not been reduced. However, Proposition 2 can now be used to deal with the leftover symbol and reduce the output alphabet size.

As mentioned previously, the existence of odd symbols in the output alphabet increases the complexity of merging symbols. In this section, solutions were given for specific cases. The following section considers a general solution for this problem.

## VII. A General Solution for "Odd" Symbols

Recall that in Lemma 2, three normal symbols were transformed into three new symbols, two normal and one leftover. In this case, the channel was upgraded. However, the output alphabet size was not reduced. Thus in Section V, an algorithm was proposed to deal with leftover symbols so as to reduce the output alphabet size while upgrading the channel. In the previous section, the existence of odd symbol in the output alphabet was considered and solutions for specific cases were provided. As previously mentioned, having odd symbols increases the complexity of upgrading the channel while simultaneously reducing the size of the output alphabet. However, this problem can be solved by considering the odd symbols as normal symbols. Recall that an odd symbol is a symbol which has zero probability for at least one input. The zero probabilities can be replaced with a very small value, i.e., $1/\varepsilon$ as $\varepsilon \to \infty$, while simultaneously keeping the sum of the symbol probabilities the same. Then all the symbols can be treated as normal, and the channel upgraded using the method given previously.

Assume that $W$ is the original channel with odd symbols in its output alphabet, and $Q'$ is the corresponding upgraded version with reduced output alphabet size. Replacing the odd symbols with normal symbols as described above results in an approximation of the original channel which is denoted by $W_{app}$. $W_{app}$ only consists of normal symbols and so can be upgraded using the method presented here. The resulted upgraded channel is denoted by $Q'_{app}$. $Q'_{app}$ is an approximation of the upgraded version of the original channel $Q'$. From this approximation, $Q'$ can be obtained according to the following asymptotic results as $\varepsilon \to \infty$

$$W = \lim_{\frac{1}{\xi} \to 0} W_{app}$$

$$Q' = \lim_{\frac{1}{\xi} \to 0} Q'_{app}$$

$$Q'_{app} \geq W_{app} \Rightarrow \lim_{\frac{1}{\xi} \to 0} Q'_{app} \geq W = \lim_{\frac{1}{\xi} \to 0} W_{app} \Rightarrow Q' \geq W$$

An example is now given to illustrate this approach to treating odd symbols. This example begins with a generalization of Lemma 2 to channels with input alphabet size $p=5$, and then continues with considering one of the output symbols as an odd symbol.

***Example:*** Let $W: X \to Y$ be a single-user channel with input alphabet size $p=5$, and let $y_1, y_2$ and $y_3$ be three normal symbols in the output alphabet $Y$.

$$y_1 = \begin{pmatrix} a_1 \\ b_1 \\ c_1 \\ d_1 \\ e_1 \end{pmatrix}, y_2 = \begin{pmatrix} a_2 \\ b_2 \\ c_2 \\ d_2 \\ e_2 \end{pmatrix}, y_3 = \begin{pmatrix} a_3 \\ b_3 \\ c_3 \\ d_3 \\ e_3 \end{pmatrix}.$$

The likelihood ratio vectors for these three symbols are

$$\lambda_1 = LR(y_1) = (\lambda_{10}, \lambda_{11}, \lambda_{12}, \lambda_{13}, \lambda_{14}) = \left(1, \frac{a_1}{b_1}, \frac{a_1}{c_1}, \frac{a_1}{d_1}, \frac{a_1}{e_1}\right)$$

$$\lambda_2 = LR(y_2) = (\lambda_{20}, \lambda_{21}, \lambda_{22}, \lambda_{23}, \lambda_{24}) = \left(1, \frac{a_2}{b_2}, \frac{a_2}{c_2}, \frac{a_2}{d_2}, \frac{a_2}{e_2}\right)$$

$$\lambda_3 = LR(y_3) = (\lambda_{30}, \lambda_{31}, \lambda_{32}, \lambda_{33}, \lambda_{34}) = \left(1, \frac{a_3}{b_3}, \frac{a_3}{c_3}, \frac{a_3}{d_3}, \frac{a_3}{e_3}\right)$$

The norm of the likelihood ratio vectors for the above symbols are then arranged in ascending order based on the norm of their likelihood ratio vectors.

Assume that $1 \leq ||\lambda_1|| \leq ||\lambda_2|| \leq ||\lambda_3|| < \infty$.

As explained in Section IV, in order to upgrade the channel, the symbol in the middle has to be divided into three new symbols, two with the same likelihood ratio vectors as the other two unchanged symbols, and the remaining leftover symbol. Denote these three new symbols as follows

$$y_{21} = \begin{pmatrix} A_1 \\ B_1 \\ C_1 \\ D_1 \\ E_1 \end{pmatrix}, y_{22} = \begin{pmatrix} A_2 \\ B_2 \\ C_2 \\ 0 \\ 0 \end{pmatrix}, y_{23} = \begin{pmatrix} A_3 \\ B_3 \\ C_3 \\ D_3 \\ E_3 \end{pmatrix}$$

The variables $A_1, A_2, A_3, B_1, B_2, B_3, C_1, C_2, C_3, D_1, D_3, E_1$ and $E_3$ can be calculated from the following equations

$$A_1 + A_2 + A_3 = a_2$$

$$B_1 + B_2 + B_3 = b_2$$

$$C_1 + C_2 + C_3 = c_2$$

$$D_1 + D_3 = d_2$$

$$E_1 + E_3 = e_2$$

$$\frac{A_1}{B_1} = \lambda_{11}, \frac{A_1}{C_1} = \lambda_{12}, \frac{A_1}{D_1} = \lambda_{13}, \frac{A_1}{E_1} = \lambda_{14}$$

$$\frac{A_3}{B_3} = \lambda_{31}, \frac{A_3}{C_3} = \lambda_{32}, \frac{A_3}{D_3} = \lambda_{33}, \frac{A_3}{E_3} = \lambda_{34}$$

These values are given by

$$A_3 = \frac{\lambda_{33}\lambda_{34}(\lambda_{14}e_2 - \lambda_{13}d_2)}{\lambda_{14}\lambda_{33} - \lambda_{13}\lambda_{34}}$$

$$A_1 = \frac{\lambda_{13}(\lambda_{14}\lambda_{33}d_2 - \lambda_{34}\lambda_{14}e_2)}{\lambda_{14}\lambda_{33} - \lambda_{13}\lambda_{34}}$$

$$B_1 = \frac{A_1}{\lambda_{11}}, C_1 = \frac{A_1}{\lambda_{12}}, D_1 = \frac{A_1}{\lambda_{13}}, E_1 = \frac{A_1}{\lambda_{14}}$$

$$B_3 = \frac{A_3}{\lambda_{31}}, C_3 = \frac{A_3}{\lambda_{32}}, D_3 = \frac{A_3}{\lambda_{33}}, E_3 = \frac{A_3}{\lambda_{34}}$$

$$A_2 = \frac{a_2(\lambda_{14}\lambda_{33} - \lambda_{13}\lambda_{34}) - \lambda_{14}\lambda_{34}e_2(\lambda_{33} - \lambda_{13}) + \lambda_{13}\lambda_{33}d_2(\lambda_{34} - \lambda_{14})}{\lambda_{14}\lambda_{33} - \lambda_{13}\lambda_{34}}$$

$$B_2 = \frac{b_2\lambda_{11}\lambda_{31}(\lambda_{14}\lambda_{33} - \lambda_{13}\lambda_{34}) - \lambda_{13}\lambda_{31}(\lambda_{14}\lambda_{33}d_2 - \lambda_{34}\lambda_{14}e_2) - \lambda_{11}\lambda_{33}\lambda_{34}(\lambda_{14}e_2 - \lambda_{13}d_2)}{\lambda_{11}\lambda_{31}(\lambda_{14}\lambda_{33} - \lambda_{13}\lambda_{34})}$$

$$C_2 = \frac{c_2\lambda_{12}\lambda_{32}(\lambda_{14}\lambda_{33} - \lambda_{13}\lambda_{34}) - \lambda_{32}\lambda_{13}(\lambda_{14}\lambda_{33}d_2 - \lambda_{34}\lambda_{14}e_2) - \lambda_{12}\lambda_{33}\lambda_{34}(\lambda_{14}e_2 - \lambda_{13}d_2)}{\lambda_{12}\lambda_{32}(\lambda_{14}\lambda_{33} - \lambda_{13}\lambda_{34})}$$

The new symbols are merged with the old remaining symbols and this way, the output alphabet of the upgraded channel is constructed.

$$z_1 = y_{21} \oplus y_1 = \begin{pmatrix} a_1 + A_1 \\ b_1 + B_1 \\ c_1 + C_1 \\ d_1 + D_1 \\ e_1 + E_1 \end{pmatrix}, z_3 = y_{23} \oplus y_3 = \begin{pmatrix} a_3 + A_3 \\ b_3 + B_3 \\ c_3 + C_3 \\ d_3 + D_3 \\ e_3 + E_3 \end{pmatrix}, z_2 = y_{22} = \begin{pmatrix} A_2 \\ B_2 \\ C_2 \\ 0 \\ 0 \end{pmatrix}.$$

Now, assume that $y_3$ is an odd symbol. To better illustrate the concept, consider the worst case odd symbol for an input alphabet size $p = 5$. Note that this odd case cannot be solved using the approach in the previous section. This odd symbol is transformed to a normal symbol as follows

$$y_3 = \begin{pmatrix} a_3 \\ 0 \\ 0 \\ 0 \\ e_3 \end{pmatrix} \Rightarrow y_{3_{app}} = \begin{pmatrix} a_3 \\ \frac{1}{3\xi} \\ \frac{1}{3\xi} \\ \frac{1}{3\xi} \\ e_3 - \frac{1}{\xi} \end{pmatrix} \quad , \quad \frac{1}{\xi} \to 0 \text{ as } \xi \to \infty$$

Substituting $y_{3_{app}}$ into the expressions for an input alphabet of size $p = 5$ gives the following values

$$\lim_{\xi \to \infty} A_1 = A_{1_{(Odd)}} = \frac{a_3(\lambda_{14} e_2 - \lambda_{13} d_2)}{\lambda_{14} e_3}$$

$$\lim_{\xi \to \infty} B_1 = B_{1_{(Odd)}} = \frac{a_3(\lambda_{14} e_2 - \lambda_{13} d_2)}{\lambda_{11} \lambda_{14} e_3}$$

$$\lim_{\xi \to \infty} C_1 = C_{1_{(Odd)}} = \frac{a_3(\lambda_{14} e_2 - \lambda_{13} d_2)}{\lambda_{12} \lambda_{14} e_3}$$

$$\lim_{\xi \to \infty} D_1 = D_{1_{(Odd)}} = \frac{a_3(\lambda_{14} e_2 - \lambda_{13} d_2)}{\lambda_{13} \lambda_{14} e_3}$$

$$\lim_{\xi \to \infty} E_1 = E_{1_{(Odd)}} = \frac{a_3(\lambda_{14} e_2 - \lambda_{13} d_2)}{\lambda_{14}^2 e_3}$$

$$\lim_{\xi \to \infty} A_3 = A_{3_{(Odd)}} = \frac{a_3(\lambda_{14} e_2 - \lambda_{13} d_2)}{\lambda_{14} e_3} = A_{1_{(Odd)}}$$

$$\lim_{\xi \to \infty} B_3 = B_{3(Odd)} = 0 \ , \ \lim_{\xi \to \infty} C_3 = C_{3(Odd)} = 0 \ , \ \lim_{\xi \to \infty} D_3 = D_{3(Odd)} = 0$$

$$\lim_{\xi \to \infty} E_3 = E_{3(Odd)} = \frac{\lambda_{14} e_2 - \lambda_{13} d_2}{\lambda_{14}}$$

$$\lim_{\xi \to \infty} A_2 = A_{2(Odd)} = a_2 - \frac{2 a_3 (\lambda_{14} e_2 - \lambda_{13} d_2)}{\lambda_{14} e_3}$$

$$\lim_{\xi \to \infty} B_2 = B_{2(Odd)} = b_2 - \frac{a_3 (\lambda_{14} e_2 - \lambda_{13} d_2)}{\lambda_{11} \lambda_{14} e_3}$$

$$\lim_{\xi \to \infty} C_2 = C_{2(Odd)} = c_2 - \frac{a_3 (\lambda_{14} e_2 - \lambda_{13} d_2)}{\lambda_{12} \lambda_{14} e_3}$$

The next step is to merge the new symbols with the previous unchanged symbols which results in the following symbols in the output alphabet of the upgraded channel

$$Q'_{app}: \ z_1 = \begin{pmatrix} a_1 + A_{1(Odd)} \\ b_1 + B_{1(Odd)} \\ c_1 + C_{1(Odd)} \\ d_1 + D_{1(Odd)} \\ e_1 + E_{1(Odd)} \end{pmatrix} , \ z_2 = \begin{pmatrix} A_{2(Odd)} \\ B_{2(Odd)} \\ C_{2(Odd)} \\ 0 \\ 0 \end{pmatrix} , \ z_3 = \begin{pmatrix} a_3 + A_{3(Odd)} \\ \dfrac{1}{3\xi} \\ \dfrac{1}{3\xi} \\ \dfrac{1}{3\xi} \\ e_3 + E_{3(Odd)} - \dfrac{1}{\xi} \end{pmatrix}$$

The final step is to calculate $Q'$ from $Q'_{app}$.

$$Q' = \lim_{\varepsilon \to \infty} Q'_{app}: \ z_1 = \begin{pmatrix} a_1 + A_{1(Odd)} \\ b_1 + B_{1(Odd)} \\ c_1 + C_{1(Odd)} \\ d_1 + D_{1(Odd)} \\ e_1 + E_{1(Odd)} \end{pmatrix} , \ z_2 = \begin{pmatrix} A_{2(Odd)} \\ B_{2(Odd)} \\ C_{2(Odd)} \\ 0 \\ 0 \end{pmatrix} , \ z_3 = \begin{pmatrix} a_3 + A_{3(Odd)} \\ 0 \\ 0 \\ 0 \\ e_3 + E_{3(Odd)} \end{pmatrix} .$$

With this approach, odd symbols can easily be incorporated into the merging process. In particular, two normal symbols to be merged with an odd symbol can be transformed into one normal symbol, an odd symbol in the same form as the original odd symbol, and a leftover symbol which can easily be handled by the output alphabet size reduction algorithm.

Note that this is a general approach. It can be applied to any type and number of odd symbols. Recall that in the special case of Lemma 3, three symbols were merged together resulting in two symbols. However, if this general approach is applied to the specific case of Lemma 3, the result will be three symbols. Therefore, it is preferable to employ Lemma 3 for the special case of an odd symbol with zero probability for only one input. In addition, for the special case of Lemma 4, applying Lemma 4 or using the general approach results in three symbols. The only difference is the form of the leftover symbols. However, the leftover symbol after applying the general approach is more compatible with the output alphabet size reduction algorithm. Thus, the general approach is preferable to complying Lemma 4 for this specific case. For all other cases, the general solution can be employed.

## *VIII. Conclusion*

This paper considered the problem for non-binary alphabets of merging output symbols in order to reduce the output alphabet size while upgrading the channel. The main intuition behind previous results for channels with a binary alphabet was explained. Based on this, an approach was developed to upgrade channels with non-binary alphabets. Unlike the degraded approximation described in [5], the upgraded approximation problem for non-binary channels is complicated. This is mainly due to the likelihood ratios of output symbols being vectors rather than scalars. The approach presented considered three categories of output symbols: normal, odd and leftover. A solution was first presented for the case when all symbols are normal (no zero probabilities for an input). Unlike the binary case, merging symbols does not reduce the output alphabet size of non-binary channels. Therefore, an output alphabet size reduction algorithm was proposed to solve this issue. It was shown that this algorithm reduces the output alphabet size to the size of the input alphabet (when all symbols are normal). The case when odd symbols (zero probabilities for at least one input) exist in the output alphabet was then investigated. Solutions for several specific cases were provided, and then a general solution for this case was given. This general approach transforms the odd symbols to normal symbols so that the solution when all symbols are normal can be employed.

## *Acknowledgement*

The authors would like to thank Prof. Alexander Vardy for his valuable comments on a previous version of the manuscript which significantly improved the final version.